%% file: main.tex
\documentclass{article}

\usepackage{microtype}
\usepackage{graphicx}
\usepackage{subcaption}
\usepackage{booktabs} 

\usepackage{hyperref}

\usepackage[accepted]{icml2026}

\usepackage{amsmath}
\usepackage{amssymb}
\usepackage{mathtools}
\usepackage{amsthm}

\usepackage[capitalize,noabbrev]{cleveref}

\theoremstyle{plain}

\theoremstyle{definition}

\theoremstyle{remark}

\usepackage[textsize=tiny]{todonotes}

\icmltitlerunning{Beyond Model Readiness: Institutional Readiness for AI Deployment in Public Systems
}

\begin{document}

\twocolumn[
  \icmltitle{Beyond Model Readiness: \\ Institutional Readiness for AI Deployment in Public Systems
}



  \icmlsetsymbol{equal}{*}

  \begin{icmlauthorlist}
    \icmlauthor{Erika Fille Legara}{ecair,aim}
    \icmlauthor{Elmo Domino Jose}{ecair}
    \icmlauthor{Paula Joy Martinez}{ecair}
  \end{icmlauthorlist}

  \icmlaffiliation{aim}{Asian Institute of Management, Makati, Philippines}
  \icmlaffiliation{ecair}{Education Center for AI Research, Department of Education, Philippines}

  \icmlcorrespondingauthor{Erika Fille Legara}{elegara@aim.edu}

  \icmlkeywords{Machine Learning, ICML, TAIGR}

  \vskip 0.3in
]



\printAffiliationsAndNotice{}  

\begin{abstract}
Many public-sector artificial intelligence systems fail not at the point of model development, but at the point of deployment. Systems that perform well in internal testing may still stall because the receiving institution lacks the approvals, data arrangements, human oversight, operational capacity, fiscal continuity, or legal clarity needed for broader rollout. Existing responsible AI and model evaluation frameworks are valuable, but they primarily assess models, datasets, and developer-side processes, not the readiness of the institution that must use the system in practice. We introduce \emph{Institutional Alignment Readiness} (IAR), a five-dimensional framework for assessing deployment readiness in public systems. The framework is designed for resource-constrained settings, where gaps between technical viability and responsible deployment are most acute. It is grounded in two anonymized operational cases from a large public education system: an image-based anthropometric screening tool and a speech-analysis system for early learning risk identification. Both reached technically viable stages but could not advance to broader rollout for institutional rather than technical reasons. We use these cases to motivate a practical readiness framework covering institutional and operational compatibility, data ecosystem maturity, human oversight capacity, fiscal sustainability, and regulatory alignment readiness. IAR is designed to complement, not replace, established AI evaluation tools. It assesses the receiving institution rather than the artifact alone and supports staging decisions such as no-go, pilot-only, or readiness for broader deployment.
\end{abstract}

\input{01_intro}
\input{02_rrl}
\input{03_iar}
\input{04_ops}

\input{05_evals}
\input{06_claims}
\input{07_practice}
\input{08_conclusion}

\section*{Impact Statement}
This paper introduces Institutional Alignment Readiness (IAR) to address a specific and preventable failure mode in public-sector AI deployment, where systems reach institutions that lack the governance, data infrastructure, oversight capacity, or legal clarity to use them responsibly. The deployment of AI in public institutions directly affects children, patients, and welfare recipients who have little visibility into the systems shaping decisions about them and the least recourse when those systems fail or are abandoned after an under-prepared pilot. Frameworks that improve deployment decisions in these settings have direct implications for equity and accountability in public services. 

Existing responsible AI frameworks evaluate models and developer-side processes; IAR fills the gap by evaluating the receiving institution, the side of the deployment equation that current tools do not assess. It gives deployment teams a structured basis for identifying and resolving institutional gaps before systems reach the populations they affect, and its value is realized when used as a judgment tool alongside technical evaluation, legal review, and meaningful community participation in deployment decisions.

\bibliography{example_paper}
\bibliographystyle{icml2026}




\end{document}

%% file: 01_intro.tex
\section{Introduction}

A recurring failure mode in public-sector AI is that technically promising systems stall between prototype and scale. The bottleneck is rarely model quality alone. More often, it is whether the receiving institution can govern, operate, and sustain the system under its own operational constraints. While machine learning evaluation is strong at assessing artifact-level properties such as accuracy, robustness, and documentation \citep{mitchell2019, raji2020}, it remains ill-equipped to surface institutional realities that dictate real-world success.

For resource-constrained public institutions, deployment is shaped by approval pathways, data-sharing protocols, frontline staffing, fiscal cycles, school or service calendars, and the realities of operating across distributed service environments. A system that performs well in internal testing may still be delayed by unresolved access agreements, limited referral capacity (qualified staff to act on outputs), operator training burdens (frontline staff preparation), or the absence of a viable maintenance pathway (funded post-deployment retraining plan) \citep{veale2019, sambasivan2021, heeks2002}. Without aligning technical design with these institutional dependencies, even high-performing prototypes inevitably succumb to the friction of distribution service environments and limited procurement capacity.

This paper argues that these deployment frictions deserve treatment as a distinct object of evaluation. Following sociotechnical accounts of AI systems, we treat institutional context not as background but as part of what determines system behavior and impact \citep{selbst2019}. We therefore introduce \emph{Institutional Alignment Readiness} (IAR), a practical framework for assessing whether the institutional conditions surrounding a specific system are sufficient for responsible deployment. Operating alongside compliance checklists, risk registers, and technical evaluations, IAR serves strictly as a staging tool to answer one specific question: is this institution ready to deploy this system at this scope?

IAR is developed from two anonymized AI projects in a large public education system: an image-based anthropometric screening tool for early-grade learners and a speech-analysis system for early learning risk identification. Both projects were stakeholder-defined, both reached technically promising stages, and both stalled on institutional rather than technical grounds during attempts to move from prototype to broader rollout. We use these cases not as a validated predictive study but as grounded evidence that deployment decisions are shaped by factors that standard model evaluation does not surface.

We make three contributions. First, we propose a five-dimensional IAR framework covering institutional and operational compatibility, data ecosystem maturity, human oversight capacity, fiscal sustainability, and regulatory alignment readiness. Second, we show through two operational cases how these dimensions help explain why technically viable systems may remain in internal validation or limited pilot rather than moving directly to scale. Third, we argue that IAR is best understood as a complement to existing responsible AI tooling. It assesses the readiness of the receiving institution, whereas most current frameworks assess the model, dataset, or developer-facing process.

%% file: 02_rrl.tex
\section{Related Work}

\subsection{Responsible AI Frameworks and Documentation Standards}

The responsible AI literature has produced a large set of principles, checklists, and documentation tools for evaluating AI systems and their development processes. \citet{jobin2019} show broad convergence around fairness, transparency, accountability, privacy, and beneficence, while also noting the persistent difficulty of translating such principles into operational practice. Model Cards \citep{mitchell2019} and Datasheets for Datasets \citep{gebru2021} improve the documentation of model and data artifacts, making intended use, limitations, and data provenance more visible. At a broader governance level, the NIST AI Risk Management Framework \citep{nistairmf2023} and ISO/IEC 42001 \citep{iso42001} provide scaffolding for organizational oversight, risk management, and accountability processes.

These contributions are essential, but they mostly focus on the AI system itself or on the organization that develops and manages it. Regulatory frameworks such as the EU AI Act introduce legal conformity assessment obligations that partially overlap with IAR's dimensions, though as externally imposed compliance requirements rather than internal readiness tools. Our concern is adjacent but distinct: whether the institution receiving a system has the authority, staffing, workflows, data arrangements, and operational capacity required to use it responsibly in practice. IAR is intended to complement, rather than replace, existing responsible AI frameworks by making the receiving-context question explicit.

\subsection{Sociotechnical Theory and Public-Sector AI Deployment}

This paper is also shaped by sociotechnical critiques of AI, which argue that model behavior cannot be understood apart from the institutional settings in which systems are deployed \citep{lipsky1980street}. \citet{selbst2019} warn against assuming systems can be transferred across contexts without rebuilding surrounding organizational supports. \citet{raji2020} show that audit practices often often miss the failures that matter most in deployment \citep{christin2017}. \citet{veale2019} document how public-sector AI is shaped by procurement constraints, legal uncertainty, and administrative risk. \citet{sambasivan2021} show that data failures in high-stakes AI reflect upstream organizational conditions. \citet{heeks2002} describes design-reality gaps endemic to information systems deployment in resource-constrained settings. IAR operationalizes this tradition by translating the sociotechnical critique into a practical pre-deployment instrument teams can act on before broader rollout.

\subsection{AI Maturity Models}

AI maturity models assess whether an organization has the general capability to adopt AI. Recent efforts have focused specifically on the public sector, proposing design-science frameworks to guide strategic organizational growth \citep{dreyling2024}. While these macro-level instruments can be useful for strategic planning, budgeting, or long-term capability development, they are designed for organization-wide strategic assessment rather than system-specific deployment decisions. A ministry, hospital, or school system may be ``AI ready'' in a broad sense, and still lack the specific referral pathways, data-sharing mechanisms, operator training, or legal groundwork required by a given model.

IAR, therefore, operates at a different level of analysis. It is system-specific rather than organization-wide, and deployment-specific rather than capability-generic. In that respect, it is closer to a readiness screen for a concrete implementation decision than to an enterprise maturity index.

\subsection{Implementation Science and Technology Readiness Frameworks}

Existing literature on technology adoption provides partial antecedents to IAR. Implementation frameworks like the Consolidated Framework for Implementation Research (CFIR) \citep{damschroder_2009_fostering} and the WHO's MAPS toolkit assess general organizational readiness, while Technology Readiness Levels (TRLs) \citep{a2016_monitoring} evaluate the technical maturity of an artifact rather than the institutional conditions, such as data-sharing agreements or human oversight, necessary for its use. 

IAR advances beyond these prior frameworks through a twofold contribution. First, it is specifically tailored to AI, operationalizing readiness dimensions shaped by the unique properties of machine learning, such as dependence on representative training data and algorithmic opacity. Second, it is system-specific rather than organization-wide; instead of measuring general implementation readiness, IAR evaluates whether an institution is fundamentally prepared to deploy a particular AI system at a specific scope, ensuring the precise legal, infrastructural, and capacity requirements of a given model are met.


%% file: 03_iar.tex
\section{Institutional Alignment Readiness: A Formal Definition}

\begin{figure*}[t]
    \centering
    \includegraphics[width=.99\linewidth]{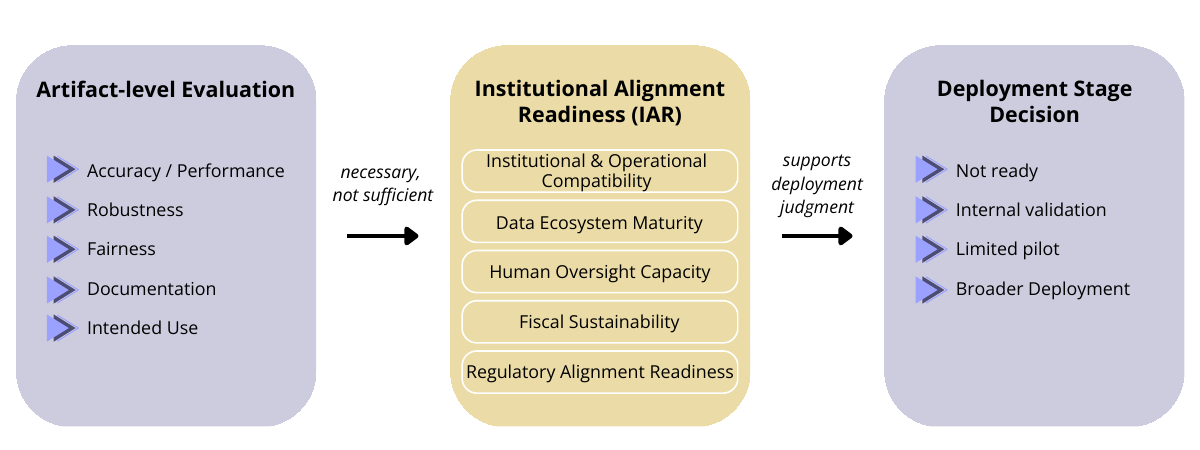}
    \caption{\textbf{From artifact-level evaluation to deployment readiness.} Existing AI evaluation tools assess whether a model or dataset is technically suitable for its intended use; IAR adds a second layer that assesses whether the receiving institution is ready to use the system responsibly at the intended deployment stage.
    \textit{Note: Although presented sequentially, IAR dimensions such as Data Ecosystem Maturity are relevant during model development and should be assessed iteratively rather than as a single pre-deployment consideration.}}
    \label{fig:iar-layers}
\end{figure*}

We define \emph{Institutional Alignment Readiness} (IAR) as a pre-deployment framework for assessing whether a target institution has the conditions needed to introduce, govern, and sustain a specific machine learning system. IAR is intended for practical deployment judgment, especially when model performance alone is insufficient to determine whether rollout is appropriate. Figure~\ref{fig:iar-layers} situates IAR within the broader deployment pipeline. Existing artifact-level evaluation remains necessary, but IAR adds a second layer focused on whether the receiving institution is ready to govern, operate, and sustain the system at the intended deployment stage.


\begin{table*}[htbp]
\centering
\caption{\textbf{IAR Dimensions: Indicators and Illustrative Failure Modes}. The dimensions are framed as practical readiness questions rather than a validated scoring instrument. The indicators and failure modes are derived from operational experience and deployment literature, and are intended to support a structured pre-deployment review. The final column makes explicit that artifact-level readiness is not sufficient for deployment: even when a system appears technically promising, rollout may still stall, narrow, or fail because the receiving institution's readiness.}
\label{tab:iar_dimensions}
\small
\begin{tabular}{|p{3.1cm}|p{5.6cm}|p{5.6cm}|}
\hline
\textbf{IAR Dimension} & \textbf{Observable Indicators} & \textbf{Illustrative Failure Mode} \\
\hline
Institutional and Operational Compatibility &
Documented approval pathway; defined go/pilot/no-go authority; audit trail capacity; workflow fit with frontline practice; operator training plan; logistics and rollout feasibility across sites; timing fit with service calendar &
System is technically ready but cannot be rolled out because approvals are unresolved, workflows are misaligned, operators are unprepared, or deployment windows are missed \\
\hline
Data Ecosystem Maturity &
Dataset representativeness for the target population; data access and sharing protocols; annotation capacity; retention and deletion policy; evidence that required data can be collected at the needed scale &
Model appears promising in development, but broader deployment stalls because target-population data are missing, siloed, or too slow to obtain \\
\hline
Human Oversight Capacity &
Availability of qualified reviewers; clear override authority; referral pathways; anti-stigma protocols; staffing continuity; mechanisms for non-technical review and challenge &
Human-in-the-loop review becomes nominal, edge cases are not escalated, or harmful outputs reach users without qualified intervention \\
\hline
Fiscal Sustainability &
Budget clarity beyond the pilot phase; maintenance and retraining plan; infrastructure and operating cost estimates; continuity plan across leadership or funding transitions &
A system works in pilot conditions but cannot be maintained, retrained, or expanded once initial project support ends \\
\hline
Regulatory Alignment Readiness &
Privacy compliance; legal basis for collection and sharing; ethics review where applicable; consent and notification procedures; contestability or redress pathway; clarity under jurisdiction-specific rules &
Deployment is delayed, narrowed, or suspended because legal classification, consent, or cross-unit data use is unresolved \\
\hline
\end{tabular}
\vspace{4pt}
\end{table*}

Importantly, IAR addresses the readiness of the \textit{receiving institution}, not the totality of deployment readiness. A system's broader deployability may also depend on artifact-level properties and the provider's or delivery team's readiness. Our claim here is narrower: even when a model clears standard evaluation thresholds, deployment may still stall because the institution expected to use it lacks the conditions for responsible use.

IAR is:
\begin{itemize}
    \item \textbf{System-specific.} It is assessed for a particular deployment, not for an institution in the abstract.
    \item \textbf{Pre-deployment.} It is most useful before broader rollout, when a team still has the option to pause, redesign, narrow scope, or defer scale-up.
    \item \textbf{Multi-dimensional.} It treats readiness as distributed across several types of institutional constraint rather than as a single score.
    \item \textbf{Operationally oriented.} It is designed to support decisions such as \emph{no-go}, \emph{pilot only}, or \emph{ready for broader deployment}, rather than providing a universal maturity ranking or a macro-level capability index \citep{dreyling2024}.
\end{itemize}

The framework comprises five dimensions, each corresponding to a distinct institutional condition required to govern, operate, and sustain a deployed system. Table~\ref{tab:iar_dimensions} summarizes the five IAR dimensions, together with illustrative indicators and failure modes drawn from the cases and the deployment literature. This matters because in public-sector deployment, institutional constraints rarely appear in isolation. Data access is tied to approvals and protocols; oversight capacity depends on real staffing and referral pathways; rollout feasibility is shaped by service calendars, geography, and operational logistics.

The five dimensions were derived by asking, across both cases, what concrete institutional conditions, if missing, would preclude the responsible deployment of a technically viable system. Each names a distinct type of dependency that recurred as a deployment constraint and maps to a recognizable gap in existing artifact-level evaluation (Table~\ref{tab:evaluation_gap}). We do not claim the five are exhaustive: conditions such as community trust, political authorization, or infrastructure interoperability may be relevant in other contexts, and we treat their absence here as a boundary condition rather than a claim about their relevance. The dimensions also reflect a deliberate prioritization of assessability. That is, we include only conditions that a cross-functional deployment team can realistically assess and act upon before broader rollout. Abstract conditions such as public trust in AI or political will for digital governance reform matter, but they are not immediately actionable at the project level. What we do claim is that each of the five dimensions included is independently necessary, wherein a system that satisfies all five is meaningfully better placed for responsible deployment than one that does not, regardless of what a more complete framework might eventually add.

While Table~\ref{tab:iar_dimensions} presents these dimensions categorically, they are practically linked by threshold and prerequisite logic. For instance, Regulatory Alignment Readiness often acts as a partial prerequisite for Data Ecosystem Maturity, as observed in Case A, the legal basis, approvals, and cross-site coordination for sharing health-proximate data had to be explicitly established before representative data collection could proceed. Similarly, regulatory and data constraints directly sequence Human Oversight Capacity. In Case B, formalizing ethics requirements, consent procedures, and data retention limits dictated which frontline staff could legally access risk scores, which in turn shaped the formal referral pathways required for human-in-the-loop oversight.

Framing readiness in multi-dimensional terms helps capture that deployment is often constrained by the interaction of several concrete institutional conditions rather than by a single abstract variable.

We intentionally do not treat IAR as a formal thresholded score in this paper. In practice, some deficits are blocking (deployment must not proceed), others are scoping (limit to pilot), and others are monitoring (proceed with active tracking). The cases do not justify strong claims about universal cutoffs, dimensional weights, or non-compensation rules. Some deficits may be absolute blockers for certain systems, while others may justify limiting deployment scope rather than stopping it entirely. For example, unresolved data-sharing legality may be a hard stop, whereas limited operator training capacity may support a pilot-only decision with tighter safeguards.

For this reason, we use IAR as a staging framework. A system may be \emph{not ready}, \emph{ready for internal validation}, \emph{ready for limited pilot}, or \emph{ready for broader deployment}, depending on which dimensions are sufficiently developed for the deployment stage under consideration. This lifecycle orientation better matches how public-sector AI systems are actually advanced in practice, where decisions are often incremental and conditional rather than binary. Because IAR does not impose universal thresholds, it is applicable across institutional contexts and system types. What transfers is the structured question, not a fixed scoring rule.

%% file: 04_ops.tex
\section{Operational Context: Two Deployments}

The cases in this paper come from two anonymized machine learning projects within a large national public education system serving learners across geographically dispersed schools. Both projects emerged from stakeholder-defined problems rather than a speculative technology push. Both also proceeded under an internal governance process that required project documentation, model documentation, impact assessment, and authorization before broader rollout. We use the cases not as proof of a validated theory, but as grounded examples of how institutional conditions shape deployment decisions.

\begin{figure*}[t]
    \centering
    \includegraphics[width=.95\linewidth]{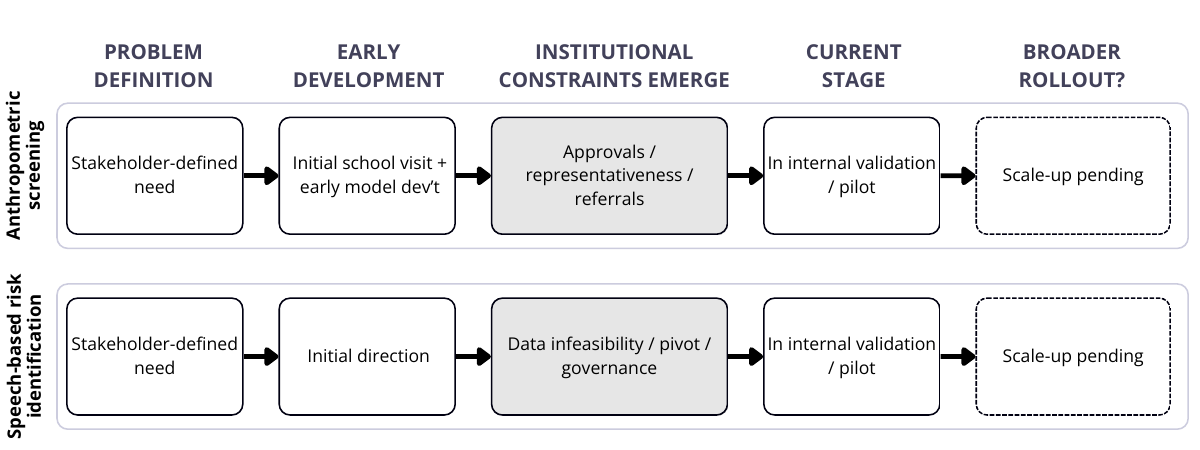}
    \caption{\textbf{Deployment trajectories of the two operational cases}. Both projects followed a similar arc from stakeholder-defined need to early development, the emergence of institutional constraints, and a current stage of internal validation or pilot. In the anthropometric screening system, the main bottlenecks concerned approvals, data representativeness, and referral capacity; in the speech-based risk identification system, data infeasibility forced an early pivot, followed by governance and coordination constraints. Broader rollout remained conditional on resolving these institutional readiness factors. Dashed boxes indicate phases contingent on those outstanding conditions.}
    \label{fig:trajectories}
\end{figure*}

\subsection{Case A: Image-Based Anthropometric Screening}
Case A uses computer vision to estimate anthropometric indicators from standard photographs of early-grade learners. The intended use is screening rather than diagnosis; that is, outputs are meant to support early identification and referral, not replace clinical judgment.

This project reached a development-ready stage quickly. An initial school visit enabled data collection and early model development within the first two months of the project. However, attempts to move toward broader deployment revealed that technical viability was only one part of the problem. Expanding data collection to additional schools took substantially longer---more than six additional months---because approval, coordination, and access had to be negotiated across multiple sites. This delay mattered not only for representativeness, but also for timing: the school calendar constrained when new data could be collected and when deployment decisions could realistically be acted upon \citep{tyack1995tinkering}.

At the time of writing, the system is best characterized as being between internal validation and limited pilot. It has been used across roughly a dozen schools, but broader scaling depends on the quality of newly collected data, the institution's ability to support more geographically diverse rollout, and the practical feasibility of deployment at the start of a new school year. Three readiness issues were especially salient: representativeness of training data (Data Ecosystem Maturity), availability of referral pathways (Human Oversight Capacity), and legal and organizational conditions for sharing health-proximate student data across units (Regulatory Alignment Readiness).

\subsection{Case B: Speech Analysis for Early Learning Risk Identification}
Case B analyzes learners' oral reading and speech samples to identify signals associated with early learning risk. As with Case A, the intended use is supportive and referral-oriented rather than diagnostic.

This project also illustrates how institutional readiness can dominate the trajectory of an AI system. The effort did not begin in its final form; it emerged after an earlier project direction proved infeasible because the required data were not available. The pivot itself is instructive; even before deployment, data feasibility acted as a decisive institutional constraint. Once the project was re-scoped, stakeholder alignment remained a central challenge. The issue was not lack of demand. Stakeholders had helped define the problem, but it still took substantial coordination to translate that demand into an operationally governable system. In resource-constrained public institutions, this sequence, where data absence forces a pivot before deployment questions even arise, is a recurrent baseline condition in our experience.

At the time of writing, this system is likewise in a limited pilot phase. Before broader deployment, internal reviewers required a set of conditions that mapped closely onto the readiness dimensions proposed in this paper. These included restrictions on who could view risk scores, validation against existing screening practice, protocols to avoid stigmatizing labels, formal referral pathways, ethics and consent requirements, data retention limits, monitoring procedures, and continuity planning across leadership changes. What matters analytically is not that this list is exhaustive or universal, but that the gating conditions were largely institutional rather than technical.

\subsection{What the Cases Show}
Taken together, the cases support a modest but important conclusion. Figure~\ref{fig:trajectories} summarizes their trajectories and highlights the institutional bottlenecks that shaped whether each system remained in validation, moved to limited pilot, or was considered for broader rollout. In both projects, the central deployment question was not simply whether the model ``worked.'' Rather, it was whether the surrounding institution could collect the right data, route outputs to qualified people, fit the system into real workflows, and sustain the deployment over time. In one case, a model-ready system was intentionally held back from faster rollout because the training dataset was still too narrow in its coverage of target-population contexts. In the other, the project trajectory itself was shaped by the availability of data, the need for multi-stakeholder protocol, and the requirement that outputs be embedded in a referral process rather than treated as standalone predictions.

These are precisely the kinds of considerations that motivated IAR. The cases establish relevance in one national education context and, by themselves, do not validate a universal measurement instrument. Cross-sector and cross-country replication by independent teams is required before IAR can be recommended as a general-purpose instrument. They do, however, show that deployment decisions in practice are already shaped by exactly the institutional conditions IAR names.

%% file: 05_evals.tex
\section{The Evaluation Gap: What Current Frameworks Cannot See}

The two cases illustrate a shared pattern: the factors that shaped deployment were not primarily model-internal, but institutional. In both cases, technically promising systems encountered bottlenecks tied to approvals, data access, staffing, workflow fit, and deployment conditions that were not visible from artifact-level evaluation alone.

Documentation standards, risk-management frameworks, and benchmark-based evaluation are designed to assess the model, the dataset, and the developer-side process. This institution on the receiving end—its authority structures, data infrastructure, operational capacity, and fiscal continuity—falls outside their scope. When deployment stalls for institutional reasons, standard tools cannot identify the problem, let alone help resolve it.
Table~\ref{tab:evaluation_gap} maps this gap across the five IAR dimensions, contrasting what current mechanisms assess against the institutional questions they leave unresolved.

The point is not that every framework fails completely in every setting. Rather, it is that none of the common tools used in ML evaluation can, on their own, answer readiness questions such as: Are the required data accessible at scale? Are qualified reviewers available?

\begin{table*}[ht]
\centering
\caption{\textbf{The Evaluation Gap by IAR Dimension}. The table contrasts common artifact-level or developer-process evaluation mechanisms with the institutional questions they typically leave unresolved in deployment. The second and third columns show what current mechanisms are designed to assess; the final column makes explicit what they often miss in practice, namely the institutional conditions that determine whether a technically promising system can move from prototype or pilot to sustained use.}
\label{tab:evaluation_gap}
\small
\begin{tabular}{|p{3.0cm}|p{3.0cm}|p{3.0cm}|p{6.0cm}|}
\hline
\textbf{IAR Dimension} & \textbf{Examples of Current Mechanisms} & \textbf{What They Primarily Assess} & \textbf{What They Typically Miss in Deployment} \\
\hline
Institutional and Operational Compatibility &
Model Cards \citep{mitchell2019}; NIST AI RMF governance processes &
Model behavior, intended use, and high-level governance recommendations &
Whether a concrete approval chain exists, whether rollout fits frontline workflows, whether operator training is feasible, and whether deployment timing and logistics are compatible with real institutional calendars \\
\hline
Data Ecosystem Maturity &
Datasheets \citep{gebru2021}; fairness metrics; robustness benchmarks &
Properties of a given dataset and model performance under distributional concerns &
Whether the necessary target-population data can actually be accessed, shared, labeled, and refreshed across the institution at the scale required for deployment \\
\hline
Human Oversight Capacity &
Human-in-the-loop design guidance; impact assessments \citep{raji2020} &
Whether human review is envisioned in the process and how it might be structured &
Whether qualified reviewers, referral pathways, override authority, and non-technical challenge mechanisms are present and sustainable in practice \\
\hline
Fiscal Sustainability &
No standard ML evaluation mechanism &
Typically outside the scope of technical evaluation &
Whether the system can survive past the pilot phase, including maintenance, retraining, operating costs, and continuity across funding or leadership transitions \\
\hline
Regulatory Alignment Readiness &
Privacy-preserving ML techniques; legal or ethics checklists &
Privacy properties at the model or data-processing level; general compliance prompts &
Whether jurisdiction-specific consent, data classification, cross-unit sharing, ethics review, and contestability requirements have been resolved for the actual deployment context \\
\hline
\end{tabular}
\vspace{4pt}
\end{table*}

The pattern across Table~\ref{tab:evaluation_gap} is structurally consistent rather than incidental. Existing mechanisms fail to surface institutional conditions not because they are poorly designed, but because they are designed for a different object of evaluation---the artifact and its development process. Fiscal Sustainability has no standard ML evaluation mechanism precisely because budgetary continuity is not a property of a model. Regulatory Alignment Readiness is addressed only at the level of data-processing techniques, not at the level of whether the specific cross-agency data-sharing required by a given deployment is legally permissible in its jurisdiction. Human Oversight Capacity is treated as a design recommendation rather than an institutional verification---existing guidance specifies that human review \emph{should} occur but provides no mechanism for confirming that the qualified humans, override authority, and referral infrastructure actually exist. These gaps cannot be closed by improving model documentation. They require a different object of assessment: the institution, not the artifact.

This distinction is consistent with prior empirical work. \citet{sambasivan2021}, for example, show that data failures in high-stakes AI deployments often reflect upstream organizational conditions rather than defects visible from the dataset alone. In similar fashion, the cases in this paper suggest that readiness failures often become legible only when one evaluates the surrounding institution: the approvals it requires, the actors who must use or contest the outputs, and the operational constraints under which rollout must occur.

IAR asks a different question than existing evaluation tools: even if a model is good enough for its intended technical function, is the institution currently in a position to deploy it responsibly at the scope being considered? The three claims that follow address whether IAR succeeds at this.


%% file: 06_claims.tex
\section{Three Claims About IAR's Value}
\paragraph{Claim 1: IAR makes deployment questions more explicit.}
IAR provides a structured way to surface readiness questions that are otherwise handled informally or too late in practice. Without such a framework, these critical issues frequently emerge only after a pilot launches, resources are committed, and reversal becomes politically difficult. These include whether the institution can access and govern the necessary data, whether qualified reviewers and referral pathways exist, whether frontline workflows and operator training can support the system, and whether there is a viable legal and fiscal basis for deployment beyond a narrow pilot. In this sense, IAR does not introduce an entirely new class of concern so much as it makes recurring deployment concerns more legible and assessable before broader rollout.

\paragraph{Claim 2: The cases suggest that these questions are decision-relevant.}
In both cases, deployment decisions were shaped by issues that standard model evaluation would not have resolved on its own. In Case A, broader rollout depended not only on model promise but on data representativeness across schools, referral capacity, and the timing constraints imposed by the school calendar. In Case B, project direction and deployment scope were shaped by data feasibility, stakeholder coordination, and the need to embed outputs within a governed referral process. These cases show that institutional readiness is practically consequential in determining whether technically viable systems remain in validation, move to limited pilot, or are considered for broader deployment.

\paragraph{Claim 3: IAR is non-redundant with existing responsible AI tooling.}
IAR addresses questions that existing frameworks are architecturally unable to answer: not whether the model is ready, but whether the institution is. Standard tools are designed specifically to assess artifacts and development processes. They do not evaluate the receiving institution and the conditions under which it can responsibly use a system. Completing a model card or a NIST RMF assessment for either operational case would have failed to surface the approval delays, referral gaps, or data-sharing constraints that determined their actual trajectories. For public-sector AI, both levels of assessment are critical.

%% file: 07_practice.tex
\section{Toward IAR as Practical Deployment Guidance}

IAR is presented here as a practice-oriented framework rather than a finalized measurement instrument. Its immediate value is as a structured deployment lens: a way for teams and institutions to ask, before broader rollout, whether the conditions for responsible use are actually in place. The point is not to produce a single abstract readiness score, but to improve the quality and consistency of deployment judgment. Several next steps would make the framework more useful in practice.

A direct extension of this work is to formalize a companion notion of provider readiness. The cases suggest that deployment can be constrained not only by the receiving institution, but also by whether the developer or delivery team can maintain the system, retrain it, support audits, respond to incidents, and manage operational transitions over time \citep{sculley2015hidden}. Sketching provider readiness dimensions, such as technical maintenance capacity, audit responsiveness, and knowledge transfer protocols, is a natural next step that would give deployment governance a complete two-sided structure: readiness of the institution that uses the system, and readiness of the team that sustains it.

Second, each IAR dimension could be translated into clearer review prompts, checklists, or rubrics suited to stage-gated decisions such as internal validation, limited pilot, and broader deployment. A cross-functional team of technical leads, compliance experts, frontline operators, and an external reviewer should conduct the IAR assessment before pilot authorization. Teams must document this assessment and revisit it at every staging decision. If deployment authority is diffuse, the IAR should directly inform a governance committee with clear escalation pathways for unresolved dimensions. The aim of such tools and oversight would not be artificial precision, but more transparent and comparable deployment decisions across projects.

Third, future work could examine how readiness expectations differ by system type and level of risk. Screening and referral systems, for example, may require stronger human oversight, clearer escalation pathways, and more robust legal review than lower-stakes administrative tools. A useful extension of IAR would therefore be to specify how deployment expectations should vary across classes of public-sector AI systems.

Fourth, cross-domain testing would help determine which readiness dimensions travel well across sectors and which require domain-specific interpretation. Education provides a useful starting point, but similar institutional questions are likely to arise in healthcare, social services, and other public-sector settings where deployment depends on organizational capacity as much as technical performance. Conditions such as referral infrastructure, data-sharing authority, and oversight staffing are configured differently across sectors, and some IAR dimensions may require recalibration or additional sub-dimensions before they can be applied with the same diagnostic precision outside the education context.

%% file: 08_conclusion.tex
\section{Conclusion}
Artificial intelligence systems in public institutions do not succeed or fail on model quality alone. In our cases, the decisive constraints were institutional: whether data could be responsibly assembled across sites, whether qualified human reviewers and referral pathways existed, whether governance authorization could be secured, and whether deployment could be supported in real operational settings. These conditions were legible before scale-up, yet they were not captured by standard model-centered evaluation. We introduce \textit{Institutional Alignment Readiness} as a practical framework for assessing these conditions systematically before deployment. For public-sector AI, the central question is not only whether a model works, but whether an institution is ready to use it responsibly.